\documentclass[prd,twocolumn,amsmath,amssymb,floatfix,superscriptaddress,nofootinbib]{revtex4-1}

\usepackage{graphicx}
\usepackage{scrextend}
\usepackage{amssymb}
\usepackage{amsmath}
\usepackage{bm}
\usepackage{color}
\usepackage{multirow}
\usepackage{cprotect}

\DeclareMathAlphabet\mathbfcal{OMS}{cmsy}{b}{n}
\definecolor{darkgreen}{RGB}{50,150,0}
\definecolor{purple}{cmyk}{0.5,1.0,0,0}

\def\edth{\;\raise1.0pt\hbox{$'$}\hskip-6pt\partial}
\def\baredth{\;\overline{\raise1.0pt\hbox{$'$}\hskip-6pt
\partial}}

\def\be{\begin{equation}}
\def\ee{\end{equation}}
\def\ben{\begin{equation} \nonumber}
\def\een{\end{equation}}
\def\ban{\begin{eqnarray*}}
\def\ean{\end{eqnarray*}}
\def\ba{\begin{eqnarray}}
\def\ea{\end{eqnarray}}
\def\({\left(}
\def\){\right)}

\definecolor{ultramarine}{rgb}{0.07, 0.04, 0.56}
\definecolor{cadmiumgreen}{rgb}{0.0, 0.42, 0.24}
\definecolor{indigo(dye)}{rgb}{0.0, 0.25, 0.42}
\usepackage[linktocpage=true]{hyperref}
\hypersetup{
colorlinks=true,
citecolor=ultramarine,
linkcolor=cadmiumgreen,
urlcolor=indigo(dye),
pdfauthor={},
pdftitle={},
pdfsubject={}
}

\begin{document}

\title{Testing consistency of $\Omega_b h^2$ in the Planck data}

\author{Pavel Motloch}
\affiliation{Canadian Institute for Theoretical Astrophysics, University of Toronto, M5S 3H8, ON, Canada}

\begin{abstract}
\noindent
We find that the cosmic microwave background temperature and polarization power spectra
measurements from Planck constrain the parameter $\Omega_bh^2$ mostly through: A) the amplitude
of Thomson scattering and B) a factor that ensures Thomson scattering does not violate
momentum conservation of the baryon-photon fluid. This allows us to obtain two
distinct but comparably strong constraints on $\Omega_b h^2$ from the Planck data alone. They are
consistent, showing robustness of the Planck $\Omega_b h^2$ constraint.  We can
alternatively rephrase these constraints as A) the change of the Thomson scattering cross
section since recombination is less than $\sim 2\%$ and B) momentum during recombination
is conserved to better than $\sim 2\%$ by Thomson scattering. Decoupling the eight various ways
in which $\Omega_b h^2$ affects the Planck data leads to $H_0$ only slightly higher than in the
standard analysis, $(68.3 \pm 1.6)\,\mathrm{km/s/Mpc}$. The overall consistency of all
$\Omega_b h^2$ constraints does not suggest any problem with the standard cosmological
model.
\end{abstract}

\maketitle

\section{Introduction}
\label{sec:intro}

Physical density of baryons $\Omega_b h^2$ is one of the parameters of the 
standard cosmological model ($\Lambda$CDM). Currently, it is best constrained
by the cosmic microwave background (CMB) temperature and polarization power spectra
measured by the Planck satellite \cite{Aghanim:2018eyx}:
\be
\label{planck}
	\Omega_b h^2 = 0.02236 \pm 0.00015 .
\ee
Measurements of primordial deuterium abundance from absorption in quasar spectra \cite{Cooke:2017cwo}
allows for competitive determinations, either
\be
\label{bbn1}
	\Omega_b h^2 = 0.02166 \pm 0.00015 \pm 0.00011
\ee
or
\be
\label{bbn2}
	\Omega_b h^2 = 0.02235 \pm 0.00016 \pm 0.00033 ,
\ee
depending on whether the value of $\mathrm{d}(\mathrm{p},\gamma)^3\mathrm{He}$ reaction
rate used to get the constraint is calculated theoretically \eqref{bbn1} or measured
\eqref{bbn2}.
The first error corresponds to uncertainty in the deuterium
abundance measurement and the second to the uncertainty of the nuclear rates and other
parts of the big-bang nucleosynthesis (BBN) calculation.

While the difference between the Planck value \eqref{planck} and the BBN value
\eqref{bbn1} is almost reaching the 3$\sigma$ level, this could just signal a problem with
the theoretical calculation of the $\mathrm{d}(\mathrm{p},\gamma)^3\mathrm{He}$ rate or a
statistical fluctuation, especially given the less precise BBN constraint
\eqref{bbn2}.

However, this discrepancy motivated us to look deeper into how exactly do the CMB data
constrain $\Omega_b h^2$ or, equivalently, through which physical processes does $\Omega_b
h^2$ enter the calculation of the CMB power spectra. Additionally, we are interested
in finding relative constraining power of these individual physical processes.
Beyond gaining theoretical understanding,
unless the baryonic constraints are strongly dominated by a single physical process, we
will be able to derive \emph{several} constraints on $\Omega_b h^2$ using the Planck data
\emph{alone}. It is not assured a priori that these should all mutually agree. This would
allow us to check the internal consistency of the Planck
data, with the hope of shedding additional light into the aforementioned
$\Omega_b h^2$ tension between Planck and deuterium measurement \eqref{bbn1}. Additionally, if a
discrepancy is found it may offer a clue on how to resolve the tension between the local
measurement of today's value of the Hubble constant, $H_0 = \(74.0 \pm 1.4\)\,
\mathrm{km/s/Mpc}$ \cite{Riess:2019cxk}, and its value inferred from CMB within the
standard cosmological model, $H_0 = \(67.3 \pm 0.6\)\,
\mathrm{km/s/Mpc}$ \cite{Aghanim:2018eyx}. In case physics beyond the standard model is responsible for this
tension, time around recombination has been singled out as the most promising place to
investigate \cite{Knox:2019rjx} and our analysis might be able to pick up signals of such
new physics.

This paper is organized as follows: In \S~\ref{sec:data} we summarize the data used and
outline the general strategy of our analysis. In \S~\ref{sec:processes} we review the
well known steps in the calculation of the CMB power spectra, list all the ways in which
$\Omega_b h^2$ enters this calculation and describe how we alter the standard computer
codes for our analysis. We present our results in \S~\ref{sec:results} and
discuss them in \S~\ref{sec:discuss}.

In this work we use the natural system of units, with the speed of light, the reduced
Planck constant, the gravitational constant and the Boltzmann constant set to unity.

\section{Data, analysis and several $\Omega_b h^2$ parameters}
\label{sec:data}

To constrain values of cosmological parameters, we use the legacy likelihoods
\verb|plik_rd12_HM_v22b_TTTEEE|, 
\verb|simall_100x143_offlike5_EE_Aplanck_B| and
\verb|commander_dx12_v3_2_29|
based on the Planck satellite measurements of the CMB temperature and polarization
power spectra \cite{Aghanim:2019ame}.

We use CosmoMC \cite{Lewis:2002ah} to obtain posterior probability distributions for the
cosmological parameters, using theoretical predictions calculated with CAMB \cite{Lewis:1999bs}. 

We use flat uninformative priors for $\Omega_c h^2$, the
physical cold dark matter (CDM) density; $n_s$, the tilt of the scalar power spectrum;
$\ln A_s$, its log amplitude at $k = 0.05\, \mathrm{Mpc}^{-1}$; $\tau_\mathrm{rei}$, the optical depth
through reionization, and $\theta_\mathrm{MC}$, the effective angular scale of the sound horizon at
recombination. We use default priors for the nuisance parameters. We run the Markov chains
until the Gelman-Rubin statistic $R-1$ drops below 0.01.

We consider eight $\Omega_b h^2$ parameters, each affecting CMB power spectra
in one of the eight different ways that are listed in \S~\ref{sec:list}. Standard analysis
would correspond to forcing all these parameters to have an identical value, we allow them to differ.
Each of these parameters is sampled with a flat prior $[0.0172, 0.0272]$ that safely
includes the independent BBN constraints \eqref{bbn1} and \eqref{bbn2}.

We assume the sum of the neutrino masses is 60 meV.
Additionally, we only consider flat cosmologies with
adiabatic initial conditions and no tensor modes.

\section{How baryons affect CMB}
\label{sec:processes}

In this section we start by reminding the reader the steps involved in the
calculation of the CMB power spectra (e.g. \cite{Dodelson:2003ft}). Then we list all the
ways in which $\Omega_b h^2$ enters the calculation and finish by describing our
implementation.

In this section we use standard symbols to denote physical quantities,
with perturbations in the synchronous gauge defined by the freely-falling dark matter
particles (see e.g.~\cite{Ma:1995ey}): $a$ is the scale factor, 
$\Omega_{b,c,\gamma,\nu}$ are fractions of today's energy
density $\rho_\mathrm{cr}$ in baryons, CDM, photons and neutrinos, $h$ and $\eta$ are
the scalar gravitational potentials, $\delta_{b,c}$ are fractional overdensities of
baryons and CDM, $v_b$ is velocity of baryons, $\Theta_{\gamma \ell}$ and $\Theta_{\nu
\ell}$ are multipole moments of the photon and massless neutrino hierarchies. 
We also label the background neutrino energy density and pressure $\bar \rho_\nu, \bar
p_\nu$, the neutrino energy density perturbation $\delta \rho_\nu$, the neutrino heat flux $q_\nu$ and
the neutrino shear stress $\sigma_\nu$.
As usual, we work with Fourier space variables.

In all equations, dot represents a derivative with respect to the conformal time.

\subsection{Steps in the CMB power spectra calculation}

As is well known, calculation of the CMB power spectra proceeds in several steps. First,
background quantities such as time dependence of the scale factor and free electron
fraction are calculated. Because of the small amplitude of
perturbations around the homogeneous and isotropic background, it is sufficient to focus
on linear perturbations, with distinct Fourier modes decoupled.
For each wavenumber $\vec k$ it is necessary to solve a set of ordinary differential equations.
Because of the assumed isotropy, only the amplitude of the wavenumber $k$ plays a role.
Once solutions for a representative set of wavenumbers are known, they are used to
calculate sources of the observed temperature and polarization fields
\cite{Seljak:1996is}. Starting with power spectra of the initial curvature perturbations
and integrating over $k$, one gets the unlensed CMB power spectra. These are then converted
into the lensed power spectra following \cite{Challinor:2005jy}.

\subsection{Effects of $\Omega_b h^2$}
\label{sec:list}

By going through the details of the CMB power spectra calculation summarized above and
checking with the source code of CAMB, we have found eight ways in which
$\Omega_b h^2$ enters the calculation of the CMB power spectra and these are listed below.

This separation is somewhat arbitrary --- for example $\Omega_b^\mathrm{eta}$ and
$\Omega_b^\mathrm{h}$ do not correspond to any gauge-invariant quantity --- but is
useful for the purpose of checking consistency of Planck data.

\subsubsection{Background expansion}

Baryons enter the time dependence of the scale factor through the Friedmann equation,
\be
\label{exp}
	\frac{\dot a}{a^2} = H_0 \sqrt{\frac{\Omega_b}{a^3} + \frac{\Omega_c}{a^3} +
	\frac{\Omega_\gamma}{a^4} + \frac{\bar \rho_\nu(a)}{\rho_\mathrm{cr}} 
	+ \Omega_\Lambda} ,
\ee
where we assume a flat Universe and thus
\be
	\Omega_\Lambda = 1 - \Omega_b - \Omega_c - \Omega_\gamma - \Omega_\nu.
\ee

\subsubsection{Helium abundance}

Helium abundance as calculated in BBN is dependent on the baryon density,
$Y_p(\Omega_bh^2, \dots)$, though this dependence is rather weak, as is well known.

\subsubsection{Recombination}

The time dependence of the free electron fraction $x_e$ calculated in
recombination codes such as Recfast \cite{Seager:1999bc} depends on the amount of baryons
in the Universe. Reader can see this easily for example on the simplest model of
recombination --- Saha equation --- in which
\be
	\frac{x_e^2}{1-x_e} = \frac{1}{n_b}\(\frac{m_e T}{2\pi}\)^{3/2} e^{-\epsilon_0/T} ,
\ee
where $m_e$ is the electron mass, $T$ CMB temperature, $\epsilon_0$ hydrogen ionization
energy and $n_b \propto \Omega_b$.

\subsubsection{Evolution of metric perturbations}

Because baryons contribute to the energy density of the Universe, their perturbations
contribute to the evolution of the synchronous gauge metric potentials \cite{Ma:1995ey},
specifically
\ba
	\frac{\dot a}{a}\dot h &=& 2k^2 \eta + 3H_0^2 \(
	\frac{\Omega_b}{a}\delta_b 
	+
	\frac{\Omega_c}{a}\delta_c
	+
	\frac{\Omega_\gamma}{a^2}\Theta_{\gamma0}
	+
  \frac{a^2 \delta \rho_\nu}{\rho_\mathrm{cr}}
	\)\nonumber\\
\label{eom_eta}
	\dot \eta &=& \frac{3H_0^2}{2k}\(
	\frac{\Omega_b}{a}v_b 
	+
	\frac{\Omega_\gamma}{a^2}\Theta_{\gamma1}
	+
  \frac{a^2 \bar \rho_\nu q_\nu}{\rho_\mathrm{cr}}
	\) .
\ea
Considering the two equations separately, we count these as two distinct effects.

\subsubsection{Thomson scattering}

Another way in which the amount of baryons in the Universe affects CMB is through the amplitude of
the Thomson scattering, as manifested for example in the equation of motion for the first photon
multipole,
\be
\dot \Theta_{\gamma1} = \frac{k \Theta_{\gamma0}}{3} - \frac{2k\Theta_{\gamma2}}{3} + a
n_e \sigma_T\(\frac{4}{3}v_b - \Theta_{\gamma1}\) .
\ee
Here $n_e$ is the free electron density and $\sigma_T$ Thomson scattering cross
section. The amplitude of the scattering, $a n_e \sigma_T$, is proportional to
$\Omega_b$.

\subsubsection{Momentum conservation}

Due to different densities of baryons and photons, the interaction term in the baryonic
equation of motion
\be
	\dot v_b = -\frac{\dot a}{a}v_b + c_s^2k \delta_b + R a n_e \sigma_T\(\frac{3}{4}\Theta_{\gamma1} - v_b\)
\ee
is multiplied by $R = \frac{4\Omega_\gamma}{3\Omega_b a}$ to ensure momentum conservation,
i.e. that the total momentum of the baryon-photon fluid is not changed by mutual
interaction through Thomson scattering. As usual, $c_s$ is baryonic sound speed.

With $a n_e \sigma_T$ already discussed, $R$ brings additional dependence on the baryonic
density.

\subsubsection{Non-linear lensing}

Finally, the lensing corrections due to non-linear effects \cite{Mead:2016zqy} also
depend on $\Omega_b h^2$, although this dependence is not expected to be significant.

\subsection{Implementation}

Having found the eight ways in which $\Omega_b h^2$ enters the calculation of the CMB
power spectra, we discuss here changes to the standard computer codes --- CosmoMC and CAMB
--- that
allow us to explore how $\Omega_b h^2$ is constrained by the CMB.

In the big picture view, we want to replace a single $\Omega_b h^2$ parameter that is part
of
the standard calculation by eight parameters $\Omega_b^X h^2$, with each entering the
calculation through a different physical process. All these parameters and what they
control are listed in Table~\ref{tab:params}.

\begin{table}
\caption{List of $\Omega_b^X h^2$ parameters}
\label{tab:params}
\begin{tabular}{cc}
\hline\hline
Parameter & Affects CMB power spectra through\\
\hline
$\Omega_b^\mathrm{EXP}h^2$ & Background expansion \eqref{exp}\\
$\Omega_b^\mathrm{BBN}h^2$ & Helium abundance\\
$\Omega_b^\mathrm{REC}h^2$ & Recombination calculation of $x_e$\\
$\Omega_b^\mathrm{ETA}h^2, \Omega_b^\mathrm{H}h^2$ & Evolution equations for $\eta, h$
 \eqref{eom_eta}\\
$\Omega_b^\mathrm{THO}h^2$ & Amplitude of Thomson scattering\\
$\Omega_b^\mathrm{R}h^2$ & Momentum conservation factor $R$\\
$\Omega_b^\mathrm{HALO}h^2$ & Non-linear lensing\\
\hline\hline
\end{tabular}
\end{table}

Most of the changes to the codes are straightforward and only consist of duplicating
variables and tracking them through the calculations, but we want to comment on two
non-trivial changes: setting the initial conditions and calculating time derivatives of
shear.

Initial conditions can be obtained by expanding all variables as polynomials in conformal
time and inverse opacity $(a n_e \sigma_T)^{-1}$ \cite{CyrRacine:2010bk} and matching
the leading coefficients in the evolution equations. After generalizing the results of Appendix B
of \cite{CyrRacine:2010bk} to our setup, we find that the following changes need to be made to the initial
conditions at conformal time $\tau_\mathrm{ini}$ as implemented in CAMB:
\ba
	\delta_\gamma^\mathrm{ini} &=& -\frac{2k^2\tau_\mathrm{ini}^2}{3}\left[1 + 
	\frac
	{\omega \tau_\mathrm{ini} (\Omega_b^\mathrm{EXP} - 3 \Omega_b^\mathrm{H} - 2 \Omega_c)}
	{10(\Omega_b^\mathrm{EXP} + \Omega_c)}
	\right]\\
	\Theta_{\nu2}^\mathrm{ini} &=& \frac{4k^2\tau_\mathrm{ini}^2}{3(4R_\nu + 15)}\left[1 + 
	\frac
	{\omega \tau_\mathrm{ini} (4 R_\nu C_1 - 5 C_0)}
	{8(\Omega_b^\mathrm{EXP} + \Omega_c)(2R_\nu + 15)}
	\right], \nonumber\\
\ea
where we have introduced
\ba
	R_\nu &=& \frac{\Omega_{\nu,\mathrm{early}}}{\Omega_{\nu,\mathrm{early}} + \Omega_\gamma}
	\\
	\omega &=& \frac{H_0(\Omega_b^\mathrm{EXP} + \Omega_c)}{\sqrt{\Omega_\gamma +
	\Omega_{\nu,\mathrm{early}}}}
	\\
	C_0 &=& 6\Omega_b^\mathrm{ETA} -7\Omega_b^\mathrm{EXP} +9\Omega_b^\mathrm{H}
	-6\Omega_b^\mathrm{R} + 2\Omega_c
	\\
	C_1 &=& -2\Omega_b^\mathrm{ETA} +5\Omega_b^\mathrm{EXP} -3\Omega_b^\mathrm{H}
	+2\Omega_b^\mathrm{R} + 2\Omega_c 
\ea
and
\be
  \Omega_{\nu,\mathrm{early}} = \lim_{a \rightarrow 0}
  \frac{\bar \rho_\nu(a)a^4}{\rho_\mathrm{cr}} 
\ee
allows us to compare energy densities of photons and neutrinos early on, when the latter
are relativistic.

Because of the way CAMB implements initial conditions, all other initial conditions are
automatically correct, e.g. they are adiabatic. It is easy to see that when all
$\Omega_b^X$ are equal, we reproduce the standard result.

\begin{figure*}
\center
\includegraphics[width = 0.99 \textwidth]{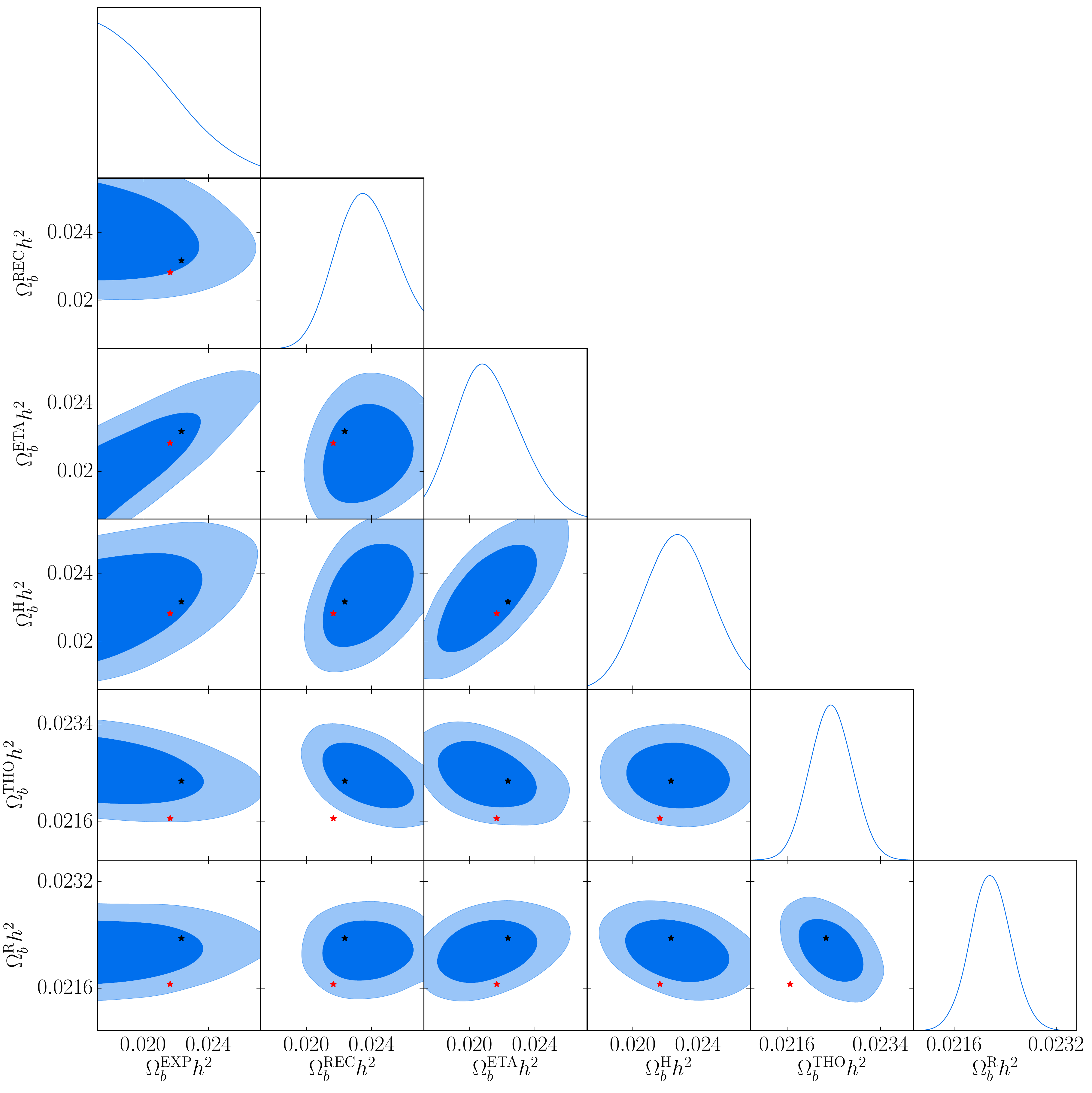}
\caption{Constraints on various $\Omega_b^X h^2$ from the Planck temperature and
polarization data (68\% and 95\% confidence limits). We omit $\Omega_b^\mathrm{BBN}h^2$
and $\Omega_b^\mathrm{HALO}h^2$ that are not constrained by the data. The
red/black stars represent mean $\Omega_bh^2$ values from the deuterium measurements
\eqref{bbn1}/\eqref{bbn2}.
}
\label{fig:b_2d}
\end{figure*}

The second nontrivial change is calculating time derivatives of shear $\sigma$,
defined as
\be
\label{sigma}
\sigma = \frac{\dot h + 6 \dot \eta}{2k} .
\ee
These derivatives are involved in second order tight coupling scheme
\cite{CyrRacine:2010bk}, and when calculating CMB sources and the gravitational lensing
potential. In CAMB, the derivatives are calculated from the Einstein equation
\be
\label{sigmadot}
	k \dot \sigma + 2 \frac{\dot a}{a}k \sigma - k^2 \eta = - 3H_0^2
	\(
  \frac{3}{2}a^2\frac{\bar \rho_\nu + \bar p_\nu}{\rho_\mathrm{cr}}\sigma_\nu
	+
	\frac{\Omega_\gamma}{a^2} \Theta_{\gamma2}
	\) ,
\ee
that is exact in general relativity.
Our altered equations of motion break \eqref{sigmadot} with terms that vanish when
all $\Omega_b^X$ are identical, which makes using \eqref{sigmadot} impractical. For
this reason we evaluate $\dot \sigma, \ddot \sigma$ directly from the definition
\eqref{sigma}, iteratively substituting equations of motion for $\dot h, \dot \eta$ and
other variables.

After finishing all the changes, we ensured that when we force all $\Omega_b^X h^2$ from
Table~\ref{tab:params} to be equal, we reproduce the standard CMB power spectra to better
than 0.01\%. Note that agreement at a machine precision level is not expected, due to the
different treatment of shear derivatives and CAMB approximations in the baryonic sector
related to neglecting most of the pressure effects of baryons
and their shear \cite{Pookkillath:2019nkn}.

\section{Results}
\label{sec:results}

In this section we present our results, first when allowing all eight $\Omega_b^X
h^2$ parameters to vary independently and then when we are more restrictive and allow only three
baryonic degrees of freedom.

\subsection{Eight baryonic parameters}

Running the CosmoMC with the Planck data when allowing all eight $\Omega_b^X h^2$ to vary,
we find that $\Omega_b^\mathrm{BBN}h^2$ and $\Omega_b^\mathrm{HALO}h^2$ are not constrained by the
data even within the very weak prior. This is in agreement with our expectations. In
Figure~\ref{fig:b_2d} we show posterior probability distributions for the remaining six
$\Omega_b^X h^2$. Of these, only
$\Omega_b^\mathrm{THO} h^2$ and $\Omega_b^\mathrm{R} h^2$ are strongly constrained; the
remaining ones are not competitive, with error bars at least four times larger. While we
find that all $\Omega_b^X h^2$ constraints are in good agreement mutually and with the BBN
constraint \eqref{bbn2}, weak tension is visible when compared with \eqref{bbn1}.

\begin{figure}
\center
\includegraphics[width = 0.49 \textwidth]{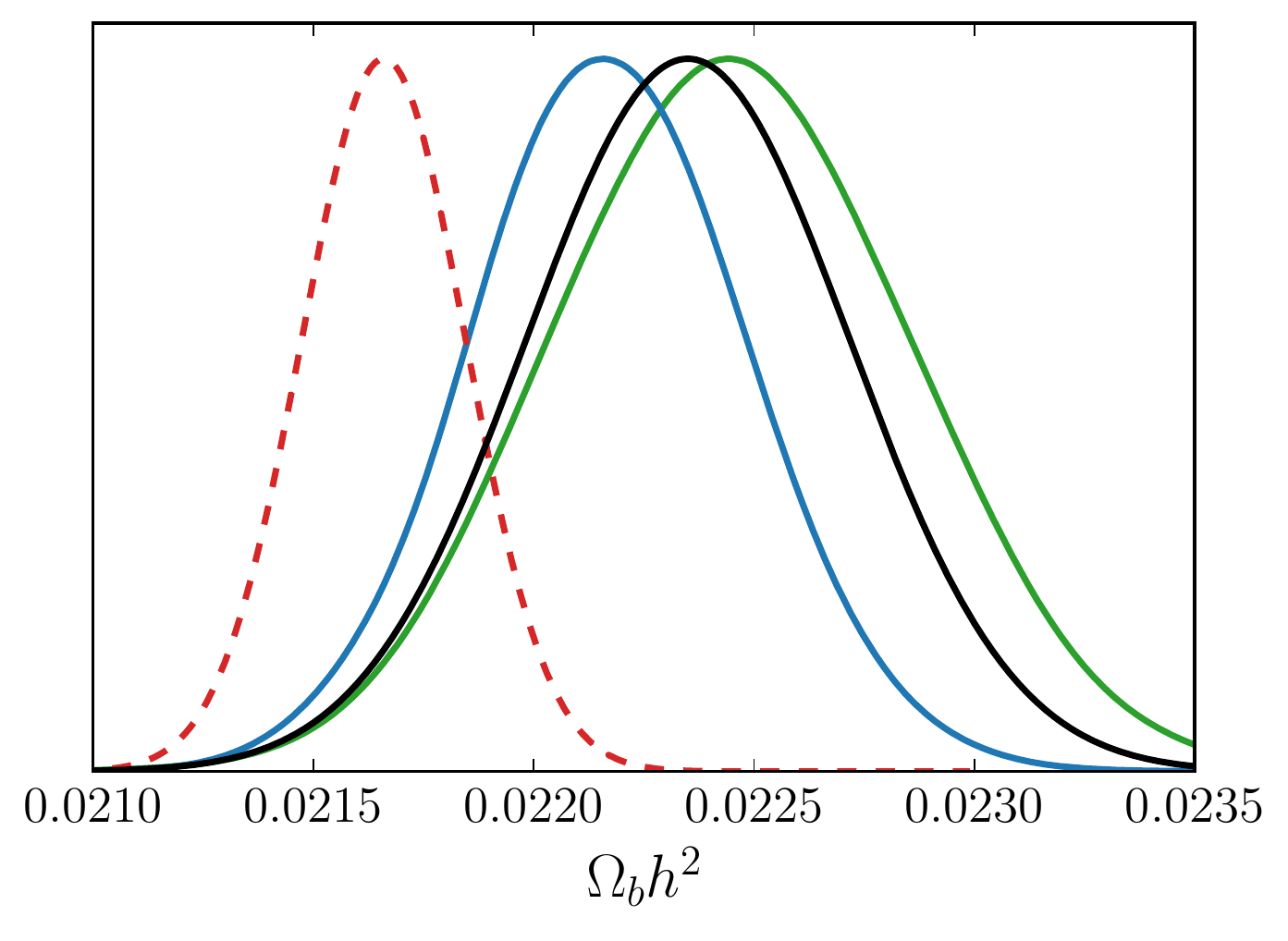}
\caption{
Constraints on $\Omega_b^\mathrm{THO}h^2$ (green) and $\Omega_b^\mathrm{R}h^2$ (blue) from
the Planck CMB power spectra when considering all eight $\Omega_b^Xh^2$ independently. For
comparison, values \eqref{bbn1} and \eqref{bbn2} derived from the deuterium abundance
measurements are shown with red dashed and black lines, assuming Gaussian posteriors and
with errors added in quadrature.
}
\label{fig:b_1d}
\end{figure}

In Figure~\ref{fig:b_1d}, we show the one-dimensional posterior probability distributions for 
the two well constrained parameters,
\ba
\Omega_b^\mathrm{THO} h^2 &=& 0.02245 \pm 0.00041
\\
\Omega_b^\mathrm{R} h^2 &=& 0.02217 \pm 0.00032
,
\ea
together with the two BBN constraints. We see that the two constraints from Planck data
are mutually consistent, which means that Planck data successfully passed our intended
consistency test.

\begin{figure}
\center
\includegraphics[width = 0.49 \textwidth]{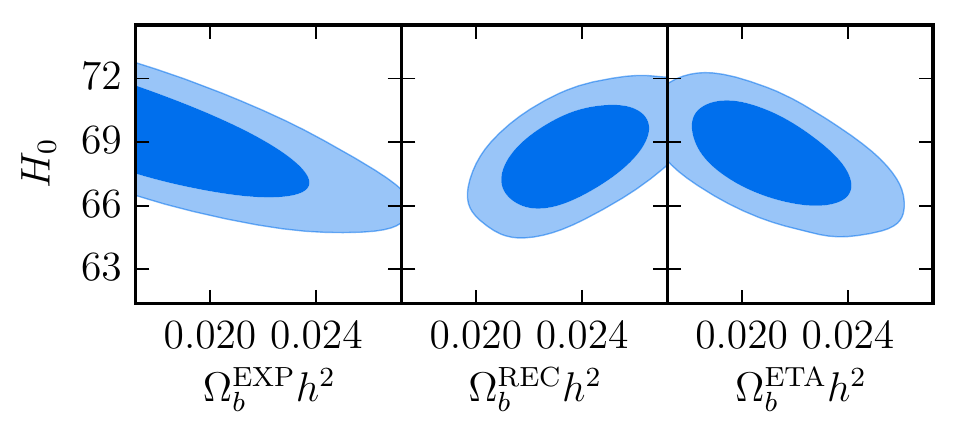}
\includegraphics[width = 0.49 \textwidth]{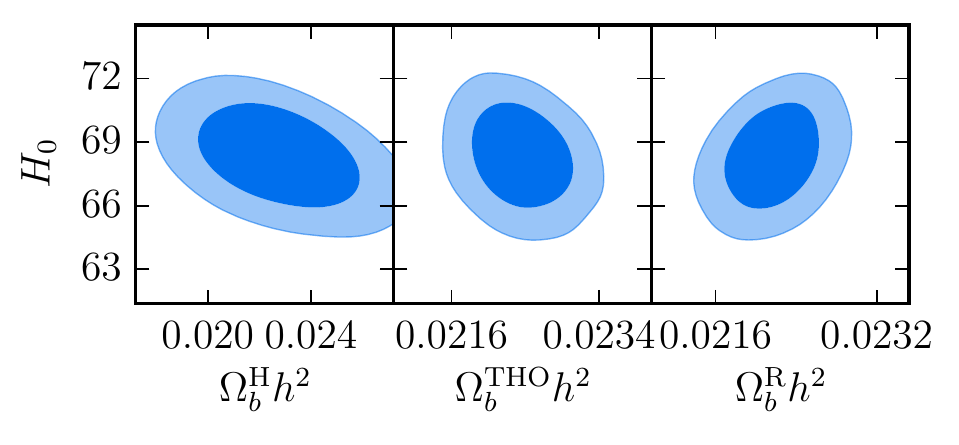}
\caption{Correlations between $H_0$ in $\mathrm{km/s/Mpc}$ and the six $\Omega_b^X h^2$
parameters constrained by Planck temperature and polarization data.
}
\label{fig:b_hubble}
\end{figure}

As another sanity check, we verified that the remaining cosmological parameters $A_s,
\tau_\mathrm{rei},
n_s, \Omega_c h^2, \theta_\mathrm{MC}$ are consistent with their standard values, although with
increased error bars. Looking particularly at constraints of $H_0$, we see its value
rise somewhat to 
\be
H_0 = \(68.3 \pm 1.6\)\, \mathrm{km/s/Mpc}.
\ee
This is driven by Planck data preferring
low $\Omega_b^\mathrm{EXP}h^2$ and $\Omega_b^\mathrm{ETA}h^2$ and high
$\Omega_b^\mathrm{REC} h^2$, see Fig.~\ref{fig:b_hubble} for correlations between $H_0$
and the six $\Omega_b^X h^2$ parameters constrained by Planck.

\subsection{Three baryonic parameters}

Given we have pinned down the two physical processes that are responsible for constraining
the density of baryons from the CMB, we can form a stronger consistency test by forcing all
the remaining $\Omega_b^X h^2$ to be equal,
\be
\label{set_equal}
\Omega_b^\mathrm{EXP} =
\Omega_b^\mathrm{BBN} =
\Omega_b^\mathrm{REC} =
\Omega_b^\mathrm{ETA} = 
\Omega_b^\mathrm{H} =
\Omega_b^\mathrm{HALO} .
\ee

\begin{figure}
\center
\includegraphics[width = 0.49 \textwidth]{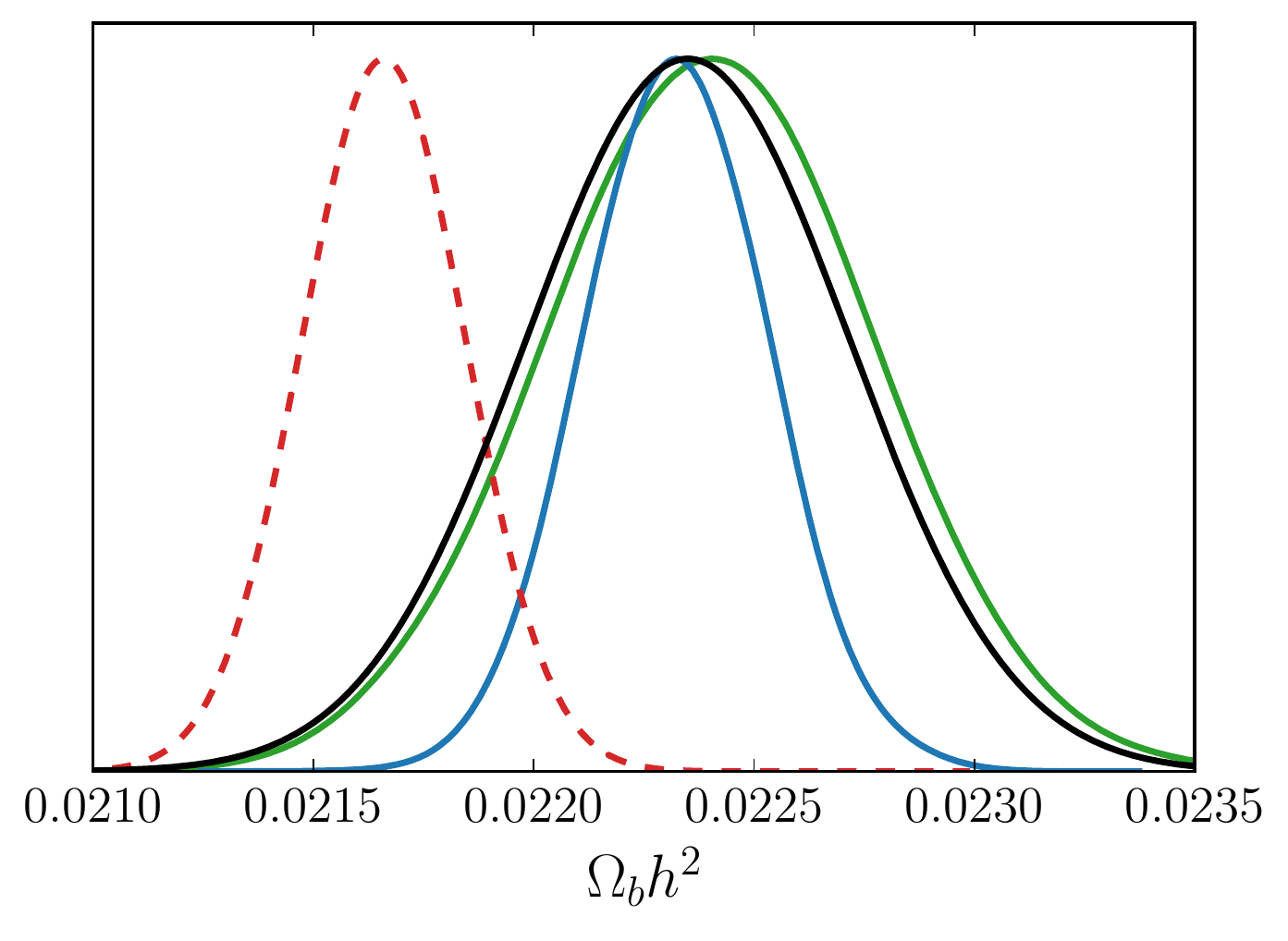}
\caption{Same as Fig.~\ref{fig:b_1d} when the six poorly constrained $\Omega_b^Xh^2$ are
forced to be identical (see \eqref{set_equal}).}
\label{fig:c_1d}
\end{figure}

Running the CosmoMC with three baryonic densities allows us to constrain
$\Omega_b^\mathrm{THO}h^2$ and $\Omega_b^\mathrm{R} h^2$ better as
\ba
\Omega_b^\mathrm{THO} h^2 &=& 0.02240 \pm 0.00037
\\
\Omega_b^\mathrm{R} h^2 &=& 0.02232 \pm 0.00021
;
\ea
these results are also shown graphically in Fig.~\ref{fig:c_1d}. Again, we find that 
the two ways in which Planck can constrain $\Omega_b h^2$ competitively are mutually
consistent and also consistent with \eqref{bbn2}. 

Regarding the constraint on the Hubble constant, it drops back to
\be
H_0 = \(67.1 \pm 0.8\)\, \mathrm{km/s/Mpc},
\ee
close to its value in the standard analysis.

\section{Discussion}
\label{sec:discuss}

We went through the calculation of the CMB temperature and polarization power spectra and
found eight distinct ways in which $\Omega_b h^2$ influences the result. By performing a
Markov chain Monte Carlo analysis with the Planck temperature and polarization data, we
found that $\Omega_b h^2$ is mostly constrained
through the amplitude of Thomson scattering and through $R$, a coefficient ensuring
that Thomson scattering conserves momentum of the baryon-photon fluid.

Given both of these constraints are mutually consistent and also consistent with the
empirical BBN constraint \eqref{bbn2}, we conclude that the proposed consistency test
passed and Planck data are internally consistent from the point of view of $\Omega_b h^2$.
This strengthens the robustness of the Planck $\Omega_b h^2$ constraints and
thus further disfavors the BBN constraint \eqref{bbn1}.

In our analysis we uncovered that Thomson scattering amplitude $an_e \sigma_T$ is
constrained at a 2\% level. Instead of using this as a constraint on $\Omega_b h^2$, we
can equivalently phrase it as $\sigma_T$ changing its value by less than $\sim 2\%$ since
the time of recombination. This is comparable to results of \cite{Hart:2019dxi}, who
constrain this change to less than 0.5\% by instead considering changes in the fine
structure constant and all natural constants dependent on it, as opposed to just
$\sigma_T$ as we do.

We also found that the parameter $R$ does not differ by more than $\sim 2\%$
from its standard model value, offering a $\sim 2\%$ test of momentum conservation in
Thomson scattering during the epoch of recombination.

When we allow the eight $\Omega_b^X h^2$ parameters to differ, we obtain a value of $H_0$
$(68.3 \pm 1.6)\,\mathrm{km/s/Mpc}$, i.e. only slightly higher than in the standard
analysis and
still far away from the local measurement \cite{Riess:2019cxk}. Restricting to three
baryonic degrees of freedom negates this preference for increased $H_0$. Overall,
the consistency of the various $\Omega_bh^2$ constraints means our analysis did not reveal
any problem with the standard cosmological model.

While details of reionization in principle also depend on the baryonic density, leveraging
this dependence into additional $\Omega_b h^2$ constraint would require a particular
physical model for the free electron fraction due to reionization. Within the
current paradigm where reionization is parameterized by a single parameter $\tau$, it is not
possible to use reionization to get an additional $\Omega_b h^2$ constraint.

Finally, any dependence of the CMB temperature and polarization power spectra on $\Omega_b
h^2$ can be traced to (generally a combination) of the eight $\Omega_b^X h^2$ discussed in
this work. For example, the linear lensing potential $C_L^{\phi\phi}$ --- and therefore
also any lensing effects -- depends on baryonic density predominantly through
$\Omega_b^\mathrm{ETA} h^2$ and $\Omega_b^\mathrm{R} h^2$.  

Only after this work was published, we learned about the existence of \cite{Chu:2004qx},
where the authors perform a somewhat similar analysis of $\Omega_b h^2$ consistency in the
CMB data. Unlike us, they use older data from WMAP and only consider two $\Omega_b h^2$
parameters. Similarly to this work, \cite{Chu:2004qx} does not find any statistically
significant deviation from $\Lambda$CDM.

\acknowledgements{
We thank an anonymous referee for useful suggestions and L. Knox for bringing
\cite{Chu:2004qx} to our attention.
}

\appendix

\bibliography{camb_baryons}

\end{document}